\documentclass[10pt,twocolumn,conference]{IEEEtran}
\usepackage{times}
\usepackage{amsbsy}
\usepackage{amssymb}
\usepackage{fontenc}
\usepackage{latexsym}
\usepackage{amsmath}
\usepackage[ansinew]{inputenc}
\usepackage[dvips ]{graphicx}
\input epsf
\usepackage{epic,eepic}
\usepackage{url}

\IEEEoverridecommandlockouts

\title{\huge Adaptive Switched Lattice Reduction-Aided Linear Detection Techniques for MIMO Systems}
\author{\textit{Keke Zu}$^\text{\textdagger}$, \textit{Rodrigo C.\ de Lamare}$^\text{\textdagger}$
\vspace*{-4em} \thanks{\footnotesize \textdagger~ Communications Research Group, Department of Electronics,
    University of York, York Y010 5DD, United Kingdom. Emails:
    \protect\url{kz511@york.ac.uk}, \protect\url{rcdl500@ohm.york.ac.uk}}}

\begin{document}
\maketitle \thispagestyle{empty} \vspace*{-1.5em}

\begin{abstract}
Lattice reduction (LR) aided multiple-input-multiple-out (MIMO) linear detection can achieve the maximum receive diversity of the maximum likelihood detection (MLD). By emloying the most commonly used  Lenstra, Lenstra, and L. Lov´asz (LLL) algorithm, an equivalent channel matrix which is shorter and nearly orthogonal is obtained. And thus the noise enhancement is greatly reduced by employing the LR-aided detection. One problem is that the LLL algorithm can not guarantee to find the optimal basis. The optimal lattice basis can be found by the Korkin and Zolotarev (KZ) reduction. However, the KZ reduction is infeasible in practice due to its high complexity. In this paper, a simple algorithm is proposed based on the complex LLL (CLLL) algorithm to approach the optimal performance while maintaining a reasonable complexity.
\end{abstract}

\section{Introduction}
    MIMO systems are fundamental for the next generation wireless networks for their great potential in improving the system capacity and performance. Two key factors that are responsible for the advantages of MIMO system are how to design the optimized signal transmission form at the transmit side and the appropriate signal detection at the receive side. The nonlinear maximum likelihood detection (MLD) can guarantee the best bit error rate (BER) performance. However, the MLD is usually impractical due to its complexity that grows exponentially with the number of constellation points and the number of transmitted streams. The linear detectors such as zero-forcing (ZF) and minimum mean square error (MMSE) receivers have the lower complexity while the simplification is the cost of sacrificing BER performance \cite{Paulraj01}.

    Recently, the lattice reduction (LR) aided detection attracted significant research efforts for it can achieve near-optimal performance with very low complexity. The LR algorithm in conjunction with MIMO detection techniques was first considered by Yao and Wornell \cite{Yao}. From the simulations of \cite{Yao}, the symbol error rate (SER) curves can parallel those of the MLD algorithm by using LR-aided detection schemes. It was proved in \cite{LrDiversity}, \cite{XiaoLM} that the LR-aided MIMO linear receivers can achieve the maximum diversity as the MLD. Hence, a great deal of interest has been devoted to exploring the application of LR in MIMO systems. The LR-aided detection schemes with respect to the MMSE criterion have been extended by Wuebben et al \cite{Wuebben}. In \cite{Windpassinger}, not only the LR-aided SU-MIMO detection but also the LR-aided SU-MIMO precoding has been investigated. LR-aided MIMO precoding for decentralized receivers was discussed in \cite{Fischer}. The quantitative error-rate analysis of LR aided detection was given in \cite{Cong02}.

  The optimal lattice basis can be found by Korkin and Zolotarev (KZ) reduction \cite{KZ}. KZ reduciton is an exponential-time lattice reduction
algorithm, hence, it is infeasible in practice due to its high complexity \cite{Closest}, \cite{Schnorr}. The most commonly used LR reduction is the polynomial-time LLL algorithm which was first proposed by Lenstra, Lenstra, and L. Lov´asz in \cite{LLL}. The LLL algorithm guarantees to find a lattice basis within a factor to the optimal one in polynomial time \cite{XiaoLM}, \cite{Closest}. The essence of the LLL algorithm is try to orthogonalize the columns of the channel matrix and reduce its size as well. The Gram-Schmidt orthogonalization (GSO) procedure and size reduction are the two core components of the LLL algorithm. Through the LLL algorithm, only a real value-based matrix can be processed which may lead to extra unnecessary complexity when the channel has large dimensions. In order to reduce the computational complexity further, the complex LLL (CLLL) algorithm was proposed in \cite{CLLL}. The overall complexity of CLLL algorithm is nearly half of the LLL algorithm without sacrificing any performance.

    In this paper, we will employ the CLLL algorithm to implement the LR-aided detection techniques for the MIMO system. As mentioned earlier, the CLLL algorithm can not guarantee to find the optimal lattice basis, hence, there will be a room left for us to improve the performance of CLLL algorithm further and maintain a low complexity at the same time. This is the motivation of this paper, and then the switched scheme is developed and proposed to improve the performance of LR-aided detection techniques.

    This paper is organized as follows. In Section II the system model and brief review of the algorithms are given. In Section III the proposed switched LR-aided MIMO detection algorithms are described in detail. Simulation results and conclusions are presented in Section IV and Section V.

{\it Notation}: Matrices and vectors are denoted by upper and lowercase boldface letters, and the transpose, inverse, pseudoinverse of a matrix $\boldsymbol B$ by $\boldsymbol B^T$, $\boldsymbol B^{-1}$, $\boldsymbol B^{\dagger}$, respectively. The $\Re$ and $\Im$ prefixes denote the real and imaginary parts. $\lceil x \rfloor$ rounds to a closest integer, while  $\lfloor x \rfloor$ to the closet integer smaller than or equal to $x$.
\section{PRELIMINARIES}
\subsection{System Model}
  In this paper, we assume an uncoded spatial multiplexing (SM) MIMO broadcast channel, where $N_T$ transmit antennas are employed at the base station (BS) and $N_R$ receive antennas are equipped at the user terminal (UT). Actually, MIMO techniques can be combined with orthogonal frequency division multiplexing (OFDM) technique. By decomposing the channel into multiple orthogonal sub-channels, the OFDM techniques can successfully transform frequency selective fading channels into flat fading channels. Hence, we assume a flat fading MIMO channel in this paper. The system model is illustrated in Fig.1, and the received signal $\boldsymbol y$ is given by:
\begin{align}
  \boldsymbol y =\boldsymbol H \boldsymbol x + \boldsymbol n,
\end{align}
where ${\boldsymbol H}\in\mathbb{C}^{N_R\times N_T}$ is the complex Gaussian channel matrix with zero mean and unit variance. We assume $\boldsymbol H$ is a full-rank MIMO channel, i.e., $\boldsymbol H$ consists of $N_T$ ($N_T\leq N_R$ ) linearly independent row vectors. ${\boldsymbol x}\in\mathbb{C}^{N_T}$ is the transmitted data vector, and ${\boldsymbol n}\in\mathbb{C}^{N_R} $ is the Gaussian noise with i.i.d. entries of zero mean and variance $\sigma_n^2$.

\begin{figure}[htp]
\begin{center}
\def\epsfsize#1#2{0.985\columnwidth}
\epsfbox{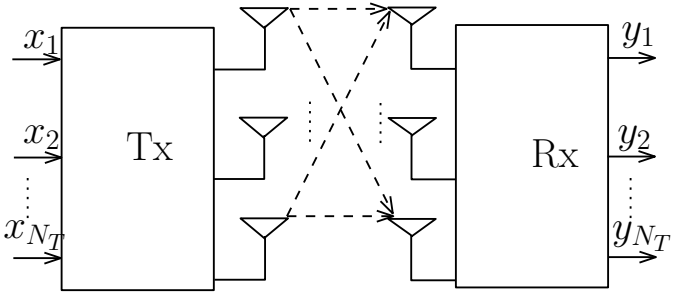} \vspace{-0.8em} \caption{\footnotesize MIMO
System Model} \label{Fig1}
\end{center}
\end{figure}

\subsection{Complex Lattice Reduction Algorithm}
A complex lattice is a set of points $L(\boldsymbol H)=\{\boldsymbol H\boldsymbol x|x_l\in\mathbb{Z}+j\mathbb{Z}\} $, where $\boldsymbol H=\{h_1,h_2,\ldots,h_{ N_T}\}$ contains the bases of the lattice $L(\boldsymbol H)$. Actually, any matrix $\boldsymbol {\tilde H}$ can generate the same lattice as $\boldsymbol H$ if and only if $\boldsymbol{\tilde H}=\boldsymbol H\boldsymbol U$, where $\boldsymbol U$ is a unimodular matrix ($det|\boldsymbol U|=1$) and all elements of $\boldsymbol U$ are Gaussian integers, i.e. $ u_{l,k} \in\mathbb{Z}+j\mathbb{Z}$.

The aim of the LR algorithm is to find a new basis $\boldsymbol {\tilde H}$ which is shorter and nearly orthogonal compared with the original matrix $\boldsymbol H$. Given the QR decomposition of $\boldsymbol H$, $\boldsymbol H=\boldsymbol Q\boldsymbol R$, where $\boldsymbol Q$ is an orthogonal matrix with unit length ($\boldsymbol Q'\boldsymbol Q=\boldsymbol I_{N_T}$) and the upper-triangular matrix $\boldsymbol R$ is a rotated and reflected representation of $\boldsymbol H$. Thus, each column vector $\boldsymbol h_k$ of $\boldsymbol H$ is given by \cite{Wuebben}
\begin{align}
\boldsymbol h_k=\sum _{l=1}^{k}r_{l,k}\boldsymbol q_l,
\end{align}
where $\boldsymbol q_l$ is the columns of $\boldsymbol Q$. If $|r_{1,k}|,..., |r_{k-1,k}|$ are close to zero, we can say that $\boldsymbol h_k$ is almost orthogonal to the space spanned by $\boldsymbol h_1,...,\boldsymbol h_{k-1}$. Similarly the QR decomposition of $ \boldsymbol{\tilde H}$ is $ \boldsymbol{\tilde H}=\boldsymbol{\tilde Q}\boldsymbol{\tilde R}$. Then, the basis for $L(\boldsymbol H)$ is CLLL-reduced if both of the following conditions are satisfied:
\begin{align}
|\Re(\tilde r_{l,k})|\leq {1\over 2}|\Re(\tilde r_{l,l})|,|\Im(\tilde r_{l,k})|\leq {1\over 2}|\Im(\tilde r_{l,l})|,1\leq l<k\leq N_T,
\end{align}
\begin{align}
\delta\|\tilde r_{k-1,k-1}\|^2 \leq \|\tilde r_{k,k}\|^2+\|\tilde r_{k-1,k}\|^2, \ 2\leq k\leq N_T,
\end{align}
where $\delta \in ({1\over2},1]$ influences the quality of the reduced basis and the computational complexity. We usually choose $\delta={3\over4}$ to achieve a trade-off between good performance and complexity \cite{LLL}. The main steps of the CLLL reduction algorithm are described clearly in \cite{CLLL}.

Obviously, $\boldsymbol {\tilde H}$ is not the unique basis for the lattice $L(\boldsymbol H)$. The lattice $L(\boldsymbol H)$ can have infinitely many different bases other than $\boldsymbol {\tilde H}$. For any unimodular matrix  $\boldsymbol U$ which satisfy $det|\boldsymbol U|=1$ and $u_{l,k} \in\mathbb{Z}+j\mathbb{Z}$, there will be a corresponding basis $\boldsymbol {\tilde H}$. Among all the bases, the optimal one can be found by KZ reduction. From \cite{Closest}, an arbitrary matrix is KZ-reduced if and only if its upper-triangular representation is KZ-reduced. For convenient, we study the upper-triangular matrix $\boldsymbol {\tilde R}$ of $\boldsymbol {\tilde H}$,
\begin{align}
{\boldsymbol {\tilde R}=\begin {bmatrix} \boldsymbol r_1\\ \boldsymbol r_2\\ \vdots\\ \boldsymbol r_{N_T}\\ \end {bmatrix}= \begin {bmatrix} r_{1,1} & r_{1,2} & ... &  r_{1,N_T} \\ 0 &  r_{2,2} & ... &  r_{2,N_T} \\  \vdots &  \vdots & \ddots &  \vdots \\ 0 & 0 & ... &  r_{N_T,N_T} \\  \end {bmatrix}},
\end{align}
the KZ reduction is recursively defined if each of the following three conditions holds:
\begin{align}
{\boldsymbol r_{N_T}\ {\rm is\ the\ shortest\ nonzero\ vector\ in}\ L(\boldsymbol H)},
\end{align}
\begin{align}
{|r_{k,N_T}| \leq {|r_{N_T,N_T}|\over 2},\ for\ k=2,...,N_T },
\end{align}
\begin{align}
{\begin {bmatrix} \boldsymbol r_{1,1} & \boldsymbol r_{1,2} & ... & \boldsymbol r_{1,N_T-1} \\ 0 & \boldsymbol r_{2,2} & ... & \boldsymbol r_{2,N_T-1} \\  \vdots &  \vdots & \ddots &  \vdots \\ 0 & 0 & ... & \boldsymbol r_{N_T-1,N_T-1} \\  \end {bmatrix}}\ {\rm is\ KZ-reduced}\,
\end{align}
 Any KZ-reduced matrix is clearly also CLLL-reduced. The optimal basis can be obtained by using KZ reduction, however, finding a KZ reduced basis would be too time-consuming and infeasible in practice. Considering the computational complexity, it is reasonable to employ the low complexity CLLL reduction algorithm to implement the lattice reduction. Consequently, there will be a performance gap between the CLLL and KZ algorithms. Regarding the performance gap, strategies are developed in section III to approach the optimal performance and offer a trade-off between performance and complexity.
\subsection{CLR Aided MIMO Detection}
The LR-reduced channel matrix $\boldsymbol {\tilde H}$ has better channel quality compared to the original channel matrix $\boldsymbol H$. Therefore, if the MIMO receivers were designed based on $\boldsymbol {\tilde H}$, a better detector performance can be achieved due to less noise enhancement increased by $\boldsymbol {\tilde H}$. The LR aided MIMO detection structure is illustrated in Fig.2,

\begin{figure}[h]
\begin{center}
\def\epsfsize#1#2{0.985\columnwidth}
\epsfbox{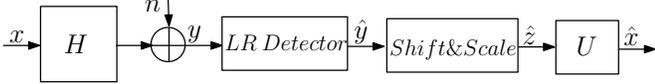} \vspace{-0.8em} \caption{\footnotesize LR Aided
MIMO Detection Structure} \label{Fig2}
\end{center}
\end{figure}
In Fig.2, $\boldsymbol x$ is the transmit M-QAM signal. The set of M-QAM constellation is given by $\mathbb S=\{\pm{1\over 2}a,\pm{3\over 2}a,...,\pm{{\sqrt M}-1\over 2}a\}$ with $\sqrt M $ representing the modulation index. The parameter $a=\sqrt{6/M-1}$ is used for normalizing the power of the transmit signals to 1. From Fig.2, the received signal is,
\begin{align}
\boldsymbol y=\boldsymbol H\boldsymbol x+\boldsymbol n=\boldsymbol H\boldsymbol U\boldsymbol U^{-1}\boldsymbol x+\boldsymbol n=\boldsymbol {\tilde H}\boldsymbol z+\boldsymbol n,
\end{align}
where $\boldsymbol {\tilde H}=\boldsymbol H\boldsymbol U$ is the CLLL-reduced channel matrix and $\boldsymbol z=\boldsymbol U^{-1}\boldsymbol x$ is the equivalent transmit signal. Similarly to the conventional MIMO detector, there are several CLR-aided MIMO detection techniques:

1) CLR-aided ZF Detector

ZF detection strategy is the simplest one, obviously, if we want to cancel out the impact of the fading channel at the receiver side, we can set the receiver filter as the pseudoinverse of $\boldsymbol {\tilde H}$ \cite{Paulraj01},
\begin{align}
 \boldsymbol {\tilde G_{ZF}}=(\boldsymbol {\tilde H^H}\boldsymbol {\tilde H})^{-1}\boldsymbol {\tilde H^H},
\end{align}
And then, the received signal is $ \boldsymbol {\breve z}=\boldsymbol z + \boldsymbol {\tilde G_{ZF}} \boldsymbol n$.

Since $\boldsymbol U$ is a unimodular matix, the statistical properties of $\boldsymbol U\boldsymbol n$ are identical to those of $\boldsymbol n$. Therefore, the estimation error covariance matrix of ZF detection is,
\begin{align}
\boldsymbol \varphi_{ZF}=E\{(\boldsymbol {\hat x}-\boldsymbol x)(\boldsymbol {\hat x}-\boldsymbol x)^H\}=\sigma_n^2\boldsymbol {\tilde G_{ZF}}\boldsymbol {\tilde G_{ZF}^H},
\end{align}
From $\boldsymbol \varphi_{ZF}$, it is clear that $ \boldsymbol {\tilde G_{ZF}}$ increased the noise interference on the signal, and thus the BER performance of the ZF detector will be degraded by $\boldsymbol {\tilde G_{ZF}}$. Since $ \boldsymbol {\tilde G_{ZF}}$ is based on the CLR-reduced channel $ \boldsymbol {\tilde H}$, it is more likely to be well conditioned compared to $ \boldsymbol G_{ZF}$ which is based on  $ \boldsymbol H$, hence the effect of noise enhancement will be moderated by $ \boldsymbol {\tilde G_{ZF}}$.

2) CLR-Aided MMSE Detector

As an improvement to reduce the effects of noise amplification caused by the ZF filter, the MMSE filter taking the noise term into account:
\begin{align}
 \boldsymbol G_{MMSE}=(\boldsymbol H^H\boldsymbol  H+\sigma_n^2\boldsymbol I_{N_T} )^{-1}\boldsymbol  H^H,
\end{align}
As shown in \cite{Wuebben02}, the MMSE detection is equal to ZF with respect to an extended system model. The extended channel matrix $\boldsymbol {\underline H} $ and the extended received signal $\boldsymbol {\underline y} $ is given by,
\begin{align}
\boldsymbol {\underline H}=\begin {bmatrix} \boldsymbol H\\ \sigma_n\boldsymbol I_{N_T}\\ \end {bmatrix}\ \rm and\ \boldsymbol {\underline y}=\begin {bmatrix} \boldsymbol y\\ \boldsymbol 0_{N_T,1}\\ \end {bmatrix},
\end{align}
From (14), it is the condition of $ \boldsymbol {\underline {\tilde H}}$ that determines the noise amplification not the condition of $ \boldsymbol {\tilde H}$ in the CLR-aided MMSE detection case. We can compute $ \boldsymbol  {\tilde G_{MMSE}}$ according to (15), however, the BER performance will be discounted for this mismatching. A better performance can be obtained by performing the CLR for the extended channel matrix $ \boldsymbol {\underline H}$, i.e.
$\boldsymbol {\underline {\tilde H}}=\boldsymbol {\underline H}\ \boldsymbol {\underline U}$, and by computing
\begin{align}
 \boldsymbol {\tilde G_{MMSE}}=(\boldsymbol {\underline {\tilde H}}^H \boldsymbol {\underline {\tilde H}})^{-1}\boldsymbol {\underline {\tilde H}}^H,
\end{align}
then, the received signal is $\boldsymbol {\breve z}= \boldsymbol {\tilde G_{MMSE}}\boldsymbol {\underline y}$.

3) CLR-Aided SIC Detector

The filers employed in Successive Interference Cancellation (SIC) detection are mainly based on the linear filters which were discussed above. The key idea of SIC detection is layer peeling, that is, the first symbol is decoded first, and then cancelling the decoded symbol in the next layer peeling, repeat this manipulation layer by layer until all the symbols are decoded from the received signal. By layer peeling, the interference caused by the already detected symbols is canceled. The SIC detection can be equivalently implemented by a QR decomposition, $\boldsymbol {\tilde H}=\boldsymbol Q\boldsymbol R$, and then computes $\boldsymbol {\tilde y}=\boldsymbol Q^H\boldsymbol y$. The detection steps is summarized bellowed,
\begin{align}
{\tilde z_{N_T}}=\bigl\lceil {\tilde y_{N_T} \over r_{N_T,N_T} }\bigr\rfloor, \\
{\tilde z_i}=\biggl\lceil  {{\tilde y_i}- \sum _{j=i+1}^{m}r_{i,j}{\tilde z_j} \over r_{i,i} }\biggr\rfloor,
\end{align}
where $i=N_T-1,...,1$. And finally, $\boldsymbol {\breve z}= [\tilde z_1,...,\tilde z_{N_T}]^T$.

After the receive filter, the estimated transmit signal $\boldsymbol {\hat x}$ can be obtained by $\boldsymbol {\hat x}=\boldsymbol U \boldsymbol {\breve z}$. However, the elements of $\boldsymbol U \boldsymbol {\breve z}$ are far from integers and serious performance loss will be caused if $\boldsymbol U \boldsymbol {\breve z}$ was quantized directly. In order to avoid the quantization error caused by rounding as much as possible, proper shifting and scaling work should be done before multiplying the received signal by $ \boldsymbol U$. The estimation of $\boldsymbol z$ is,
\begin{align}
 \boldsymbol {\hat z}=a(\boldsymbol {\hat {\bar z}}+{1\over 2}\boldsymbol U^{-1}\boldsymbol 1_{N_T}),
 \end{align}
where $\boldsymbol {\hat {\bar z}}=\lceil {1\over a}\boldsymbol {\breve z}-{1\over 2}\boldsymbol U^{-1}\boldsymbol 1_{N_T}\rfloor$. Finally the estimated transmit signal $\boldsymbol x$ can be easily obtained by $\boldsymbol {\hat x}=\boldsymbol U \boldsymbol {\hat z}$.

\section{Proposed Switched LR-Aided Detection}
The proposed algorithm is motivated by the fact that the CLLL-reduced channel matrix $\boldsymbol {\tilde H}$ is neither the optimal nor the unique basis in the lattice space. Therefore, we can build more LR-reduced candidate matrices and choose the best one during each detection stage. Observing the formulation $\boldsymbol {\tilde H}=\boldsymbol H\boldsymbol U $, it is not difficult to find that there are two ways if we want to get more LR-reduced candidates.

The first way is achieved by constructing different satisfied $\boldsymbol U$ matrix. As long as $\boldsymbol U$ always meet two requirements, (1) $det|\boldsymbol U|=1$, (2) all elements of $\boldsymbol U$ are Gaussian integers. The physical meaning of the first requirement is that we cannot change the transmit power through the LR transformation. Under the first limitation, we need to construct the different $\boldsymbol U$ matrices by the four units 1, -1, j and -j.

The second way to get more $\boldsymbol {\tilde H}$ candidates can be achieved by changing $\boldsymbol H$,  that is, we can randomly swap columns of channel matrix $\boldsymbol H$ at the receiver side before the LR transformation. Actually, interchanging the columns of $\boldsymbol H$ is equivalent to performing a linear transformation of $\boldsymbol U$.

 Comparing these two ways, it is not easy to construct different satisfied $\boldsymbol U$ matrices by the four units, while, the second way is more convenient to implement. Hence, we focus on the second scheme to generate more candidates of $\boldsymbol {\tilde H}$. The art of switched techniques are also discussed in \cite{Rodrigo}.
\subsection{The System Structure of the Proposed Algorithm}
The LR-reduced candidate channel matrices $\boldsymbol{\tilde H_s}$ are roughly orthogonal, consequently, we need to develop a metric to measure the orthogonality and choose the best one among them. The system model of the proposed randomly switched CLR-aided detection algorithm is illustrated in Fig.3.

\begin{figure}[htp]
\begin{center}
\def\epsfsize#1#2{0.985\columnwidth}
\epsfbox{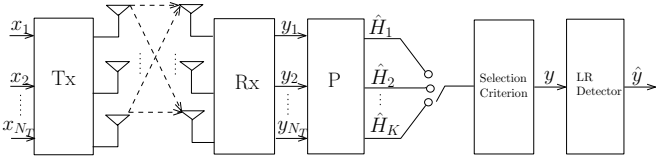} \vspace{-0.8em} \caption{\footnotesize Switched
LR Aided Detection Structure} \label{Fig.3.}
\end{center}
\end{figure}
From Fig.3 $\boldsymbol P$  denotes the randomly permuting matrix which exclude the original order, the corresponding LR-reduced channel matrix is given by,
\begin{align}
 \boldsymbol {\tilde  H_i}=(\boldsymbol H\boldsymbol P_i) \boldsymbol U_i,\ i \in (1,2,...,K), K\leq N_T!,
\end{align}
where $i$ is the $i-th$ candidate, and $K$ is the total number of candidates. Considering the complexity of the algorithm, the number of candidate matrices is limited to no more than 10 when the channel matrix has high dimensions, that is, $ K\leq N_T!\ and\ K\leq 10$.
\subsection{The Selection Criterion}
 Obviously, designing powerful selection criterion is crucial to the proposed algorithm. From matrix theory \cite{MatrixTheo}, the condition number (CN) can be used as the indication of the orthogonality of the reduced matrix. The CN is defined as,
 \begin{align}
  \kappa(\boldsymbol {\tilde  H})={\sigma_{max}\over \sigma_{min}},
  \end{align}
  where $\sigma_{max}$ and  $\sigma_{min}$ is the largest and smallest singular value of $\boldsymbol {\tilde  H}$. For the orthogonal channel matrix the value of CN is 1, there will be no noise amplification effect during the detection. The matrix with low CN is well-conditioned, while the matrix with a high CN is ill-conditioned. In \cite{Wuebben03} the CN is used to study the impact of channel matrix on the different detection techniques. It is showed by the simulation results that all the detection schemes can achieve very good performance for $\kappa(\boldsymbol H)\approx 1$. However, CN mainly reflects the worst situation of the channel matrix and thus it is only a rough measurement. Therefore, CN is not a reliable selection criterion due to the fact that CN cannot fully measure the channel performance.

 Luckily, we found a stable and reliable metric to serve as the selection criterion called orthogonality defect factor (ODF) which is defined in \cite {LrDiversity} as,
 \begin{align}
 ODF(\boldsymbol H)={(\|\boldsymbol h_1\|^2\|\boldsymbol h_2\|^2...\|\boldsymbol h_{N_T}\|^2)\over det(\boldsymbol H^H\boldsymbol H)},
\end{align}
 where $\boldsymbol h_i$s are the columns of the basis $\boldsymbol H$. Clearly, $ODF(H)\leq 1$ with equality of an orthogonal basis. Therefore, for candidates $\boldsymbol {\tilde H_1}$ and $\boldsymbol {\tilde H_2}$, we can say  $\boldsymbol {\tilde H_1}$ is better reduced than $\boldsymbol {\tilde H_2}$ if $ODF(\boldsymbol {\tilde H_1})\leq ODF(\boldsymbol {\tilde H_2})$. By using the ODF selection metric, we can choose the best candidate $\boldsymbol {\tilde H_{op}}$ from $\{\boldsymbol {\tilde H_1},\boldsymbol {\tilde H_2},...,\boldsymbol {\tilde H_K} \}$, and then the received signal can be rewritten as,
 \begin{align}
\boldsymbol y=\boldsymbol {\tilde H_{op}}\boldsymbol z+\boldsymbol n,
\end{align}
 consequently, the receive filter is calculated based on $\boldsymbol {\tilde H_{op}}$ as well. Since $\tilde H_{op}$ has better channel quality, the BER performance will be improved by the proposed algorithm.

 The complexity of the proposed algorithm is determined by the total candidate number $K$. The different trade-offs between performance and complexity can be achieved by altering $K$ depending on specific situation. For example, we can set $K=1$, and compare the orthogonality between the candidate and the original channel matrix. Then, choose the better one to compute the detection filter. Hence, the performance will be always better or equal to the conventional CLR-aided detection.

  A graceful trade-off between BER performance and computational complexity offered by the proposed algorithm which is summarized in table I. The simulation results are given in the next section.
 {\footnotesize
\begin{table}[h]
\centering{\textbf{Table I: Proposed Switched LR-Aided Algorithm}}\\
\begin{tabular}{|l|}
\hline
\\
~~~\textbf{Function $\boldsymbol H_{KLR}$=KLR($\boldsymbol H, \boldsymbol K$)} \\
~~~1: $\boldsymbol H_{LR}=CLLL(\boldsymbol H)$\\
~~~2: $\boldsymbol Tao=ODF(\boldsymbol H_{LR})$\\
~~~3: $\boldsymbol P=Rand_-Permute(K)$\\
~~~4: for i=1:K \\
~~~5: $\boldsymbol H_{LR_i}'=CLLL(\boldsymbol H\boldsymbol P_i)$\\
~~~6: $\boldsymbol Tao'_i=ODF(\boldsymbol H_{LR_i}')$\\
~~~7: end\\
~~~8: [Min~ Idx]=Min($\boldsymbol Tao'$)\\
~~~9: if $\boldsymbol Tao'_{Indx} < \boldsymbol Tao$\\
~~~10: ~~~~$\boldsymbol H_{KLR} = \boldsymbol H_{LR_{Idx}}'$\\
~~~11: \textbf{else}\\
~~~12: ~~~~$\boldsymbol H_{KLR} = \boldsymbol H_{LR}$\\
~~~13: \textbf{end}\\
~~~14: \textbf{return} $\boldsymbol H_{KLR} $\\
\hline
\end{tabular}
\end{table}
}

\section{Simulation Results}
  In this section, the linear MIMO detection techniques are given and compared with the proposed algorithms. A system with $\boldsymbol N_T=6$ transmit antennas and $\boldsymbol N_R=6$ receive antennas is considered. We assume an uncorrelated block fading channel, that is, the channel is static during each transmit packet and there is no correlation among the antennas. The number of trials used to average the curves is 1000 and the packet length is 100 symbols for the simulations. The SNR is defined as $SNR={N_T\sigma_x^2\over\sigma_n^2}$, and the $E_b/N_0$ is defined as $E_b/N_0=SNR{N_R\over N_TR_m}$ with $R_m$ is the number of information bits transmitted per channel symbol $x_l$.

Fig.4 showed the BER performance of the proposed randomly switched CLR-aided schemes with ZF detection. From Fig.4, we can find that the ZF detection has the worst BER performance compared to the other techniques since the high noise amplification without LR. The BER performance of CLR-aided ZF detection is better than the conventional ZF detection while worse than the proposed algorithm. In addition, the performance of the proposed scheme gradually improved as the candidates number $K$ increases. It is clear that the proposed scheme have better performance compared to ZF detection and LR-aided ZF detection. At the BER of $10^{-3}$, the proposed scheme with $K=10$ has more than 4.5dB gains compared to the CLR-aided ZF detection. It is worth noting that even with $K=1$, the proposed scheme has almost 2dB gains over the CLR-aided ZF linear detection.

Fig.5 illustrated the MMSE,MMSE-SIC and their corresponding proposed algorithms respectively. The MMSE based detection algorithms have better BER performance compared to their existing ZF counterparts in Fig.4. At 25dB, the BER of conventional MMSE and CLR-aided extended MMSE detection are all higher than $10^{-4}$. While, the BER of the proposed scheme achieved very good performance at 25dB, they are all below $10^{-4}$ and even approached $10^{-5}$ with $K=10$. Since there is a interference cancellation during each SIC detection layer, the improved performance of the proposed randomly switched CLR-aided MMSE-SIC is quite limited compared with the randomly switched technique with linear detection. However, the proposed algorithm still offer almost 2dB gain with $K=10$ at $10^{-4}$ compared to CLR-aided MMSE-SIC.

\begin{figure}[htp]
\begin{center}
\def\epsfsize#1#2{0.985\columnwidth}
\epsfbox{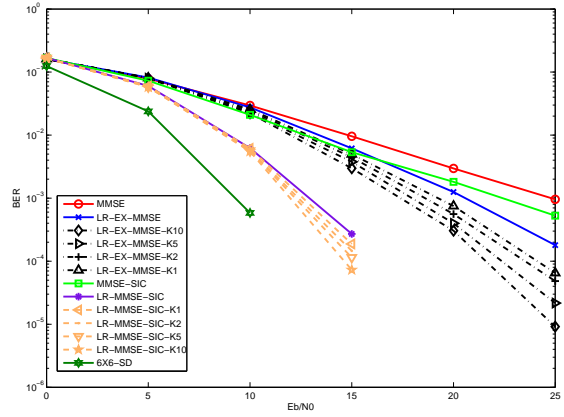} \vspace{-0.8em} \caption{\footnotesize
Switched-CLR-ZF, 6X6 MIMO, QPSK} \label{Fig.4.}
\end{center}
\end{figure}

\begin{figure}[htp]
\begin{center}
\def\epsfsize#1#2{0.985\columnwidth}
\epsfbox{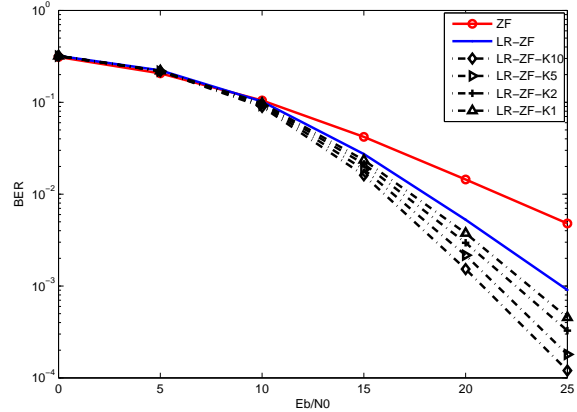} \vspace{-0.8em} \caption{\footnotesize
Switched-CLR-MMSE and MMSE-SIC, 6X6 MIMO, QPSK} \label{Fig.5.}
\end{center}
\end{figure}
Fig.6 illustrated the BER performance with 16-QAM modulation. Obviously, the BER performance of all corresponding algorithms are becoming worse compared to the QPSK modulation scenario due to the smaller signal point distance. Similarly, the proposed schemes offered better performance in the 16-QAM modulation.

\begin{figure}[htp]
\begin{center}
\def\epsfsize#1#2{0.985\columnwidth}
\epsfbox{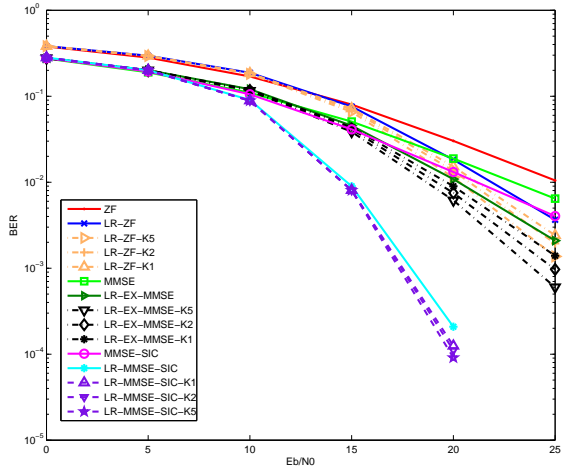} \vspace{-0.8em} \caption{\footnotesize
Switched-CLR-ZF, MMSE and MMSE-SIC, 6X6 MIMO, 16-QAM} \label{Fig.6.}
\end{center}
\end{figure}

These simulation results support the analysis that the LR algorithms based on LLL can not guarantee an optimal basis, however, we can approach the optimal performance by using the proposed scheme. Actually, if the equivalent channel is strictly orthogonal between each other, the linear detection is identical to MLD \cite {Wuebben}.
\section{conclusion}
In this paper, the randomly switched technique was proposed to improve the performance of LR-aided MIMO detection, which gracefully offered a tradeoff between the complexity and performance. Simulation results evidence that our proposed algorithms have substantial performance gains compared to the existing MIMO linear and LR-aided linear detection.  As for nonlinear SIC detection, the gain improved by the proposed algorithm is limited. How to reduce the gap between the proposed algorithm based on SIC detection and the ML or SD detection need further study.

\end{document}